\documentclass[conference]{IEEEtran}

\usepackage[T1]{fontenc} 
\usepackage{amsmath}
\usepackage[noadjust]{cite}
\usepackage{txfonts}
\usepackage{bm} 
\usepackage{graphicx}
\graphicspath{ {../../Projects/DigitalHumanities/SovietMusicMap} }
\usepackage{booktabs}

\author{\IEEEauthorblockN{Dmitry Zinoviev}
\IEEEauthorblockA{Mathematics and Computer Science Department\\
Suffolk University\\
Boston, Massachussets 02114\\
Email: dzinoviev@suffolk.edu}
}
\title{Networks of Music Groups as Success Predictors}

\begin{document}
\maketitle
\begin{abstract}
More than 4,600 non-academic music groups emerged in the USSR and post-Soviet independent nations in 1960--2015, performing in 275 genres. Some of the groups became legends and survived for decades, while others vanished and are known now only to select music history scholars. We built a network of the groups based on sharing at least one performer. We discovered that major network measures serve as reasonably accurate predictors of the groups' success. The proposed network-based success exploration and prediction methods are transferable to other areas of arts and humanities that have medium- or long-term team-based collaborations.
\end{abstract}


\section{Introduction}

Exploring and predicting the success of creative collaborations, such as scientific research teams~\cite{ghasemian2016}, teams of video game developers~\cite{vaan2015}, artistic communities~\cite{giuffre1999}, online discussion boards~\cite{preece2001}, theatric~\cite{hirschman1985} and cinematographic~\cite{hadida2010} projects, and music groups and bands~(\cite{uzzi2005,mauskapf2016}), has been an area of active research in the past two decades. As early as in 1973, Granovetter ~\cite{granovetter1973} suggested that the topology of an individual's social network has an impact on the personal success. The connection between network topology and success was also found in the networks of collaborations of individuals~(\cite{ghasemian2016,vaan2015,uzzi2005}). Such networks can be constructed, e.g., based on performers sharing, when two groups or bands are connected if there is at least one performer who participated in both teams (not necessarily at the same time).

The goal of this study is to investigate if sharing performers with other groups indeed influences the groups' eventual success, and predict the success, based on performers sharing. For the study, we chose Soviet and post-Soviet popular (non-academic) music groups. 

Several thousand non-academic music groups emerged within the borders of the former USSR in 1960--2015, performing in 275 genres and sub-genres (we use the Wikipedia genre classification), including rock, pop, disco, jazz, and folk. Some of the groups became legends and survived for decades, while others vanished and are known now only to select music history scholars and fans. The Soviet/post-Soviet popular music landscape has been mostly ignored by recent social and music history researchers---probably due to the language barrier and relative lack of popularity of Soviet/Russian music in the West. To the best of our knowledge, only one in-depth study of early Russian rock has been published in English~\cite{troitsky1988}. On a broader scale, Pilkington~\cite{pilkington2001} discusses the process of penetration of the Western popular music culture into the post-Soviet realm but fails to outline the aboriginal music landscape. On the other hand, Bright~\cite{bright1986} presents an excellent narrative overview of the non-academic music in the USSR from the early XXth century to the dusk of the Soviet Union---but with no quantitative analysis. Sparse and narrowly scoped studies of few performers~\cite{partan2007} or aspects~\cite{steinholt2003} only make the barren landscape look more barren.

In this paper, we apply modern quantitative analysis methods, including statistical analysis, social network analysis, and machine learning, to a collection of 4,600 music groups and bands. We build a network of groups, quantify groups' success, correlate it with network measures, and attempt to predict success, based solely on the network measures.

\section{Literature review}
Roy and Dowd~\cite{roy2010} proposed a sociological approach to the music industry. In particular, they looked at the interaction between individuals and groups and the possibility of collective production of music.

Attempts to explain or predict the success of creative individuals through social network analysis have been undertaken as early as in 1999, essentially immediately following the seminal paper by Krackhardt and Carley~\cite{krackhardt1998}. Giuffre~\cite{giuffre1999} conceptualized artists' careers as a trajectory through a longitudinal social network of ad hoc relationships. Her study covers the period from 1981 to 1992 and identifies three distinct career paths with differential success outcomes.

Uzzi and Spiro~\cite{uzzi2005} analyzed a network of creative artists who made Broadway musicals from 1945 to 1989. By applying statistical methods, they found that the network measures of the network of those artists affected their creativity. They discovered that effect of the network was parabolic: performance increased up to a threshold, after which point the positive effects reversed. 

The networking approach to success was further elaborated in the 2010s, as large-scale collections of cultural and artistic data, and computational resources compatible with the amounts of data became available to social science and humanities researchers. Park et al.~\cite{park2015} explored the complex network of Western classical composers and correlated network measures, such as degree distributions and centralities, with composer attributes. They also investigated the growth dynamics of the network. Essentially at the same time, a model predicting success of scientific collaborations was built based on the analysis of networks of previous collaborations of scholars, using machine learning techniques~\cite{ghasemian2016}. The only network feature that they utilized for the analysis was the degree centrality.

Mauskapf et al.~\cite{mauskapf2016} and Mauskapf and Askin\cite{mauskapf} proposed a new explanation for why certain cultural products outperform their competitors to achieve widespread popularity, based on the product position within their cultural networks. They tested how the musical features of over 25,000 songs from Billboard's Hot 100 charts structured the consumption of popular music and found that song's position in its cultural network influences its position on the charts. 

Multiple definitions of success and creativity have been proposed as well, including the number of members, number of posted messages, members satisfaction, reciprocity of communications, trustworthiness, and usability in and of a creative online community~\cite{preece2001}, the amount of critical notice~\cite{giuffre1999}, the financial and artistic performance of Broadway musicals~\cite{uzzi2005} and cinema projects~\cite{hadida2010}, or song's position in the charts~\cite{mauskapf2016}.

\section{Method}
For this study, we collected structural information about the network of Soviet and post-Soviet non-academic music groups and bands, as well as success-related metrics.

\subsection{Network Construction}

The most complete, trusted, and well-organized (though far from being perfect) source of data on the groups and bands in the USSR and the post-Soviet independent states is Wikipedia. We collected the raw data from Wikipedia in August--September 2015. The downloaded dataset consists of 4,367 pages in seven primary languages: Russian (ru, 2,174), Estonian (et, 894), Ukrainian (uk, 724), Latvian (lv, 176), Lithuanian (lt, 172), Belarusian (by, by-x-old, and by-tarask, 154), and Moldavian (ro, 27). Another 115 languages were used for translated versions of the pages. Overall, the pages cover 4,560 groups and 16,329 individual performers (on average, 3.6 performers per group; at least 3,600 of them participated in more than one project). We noticed that the Trans-Caucasian and Central Asian states (Armenia, Azerbaijan, Georgia, Kazakhstan, Kyrgyzstan, Tajikistan, Turkmenistan, and Uzbekistan) did not have a critical mass of popular music groups or those groups were not present on Wikipedia, even in the national language segments. About 80\% of performers and groups do not have their own Wikipedia pages but are mentioned in other pages. 

Unlike their English-language counterpart, the Wikipedias in the languages of our interest do not use uniform templates for music groups and performers. There is no directory of all music groups, and the directories of groups by genre are incomplete.  These irregularities did not allow us to develop simple software to extract the essential information and automate the download; we ended up downloading the pages by hand.

\begin{figure}[!t]
\centering
\includegraphics[width=\columnwidth]{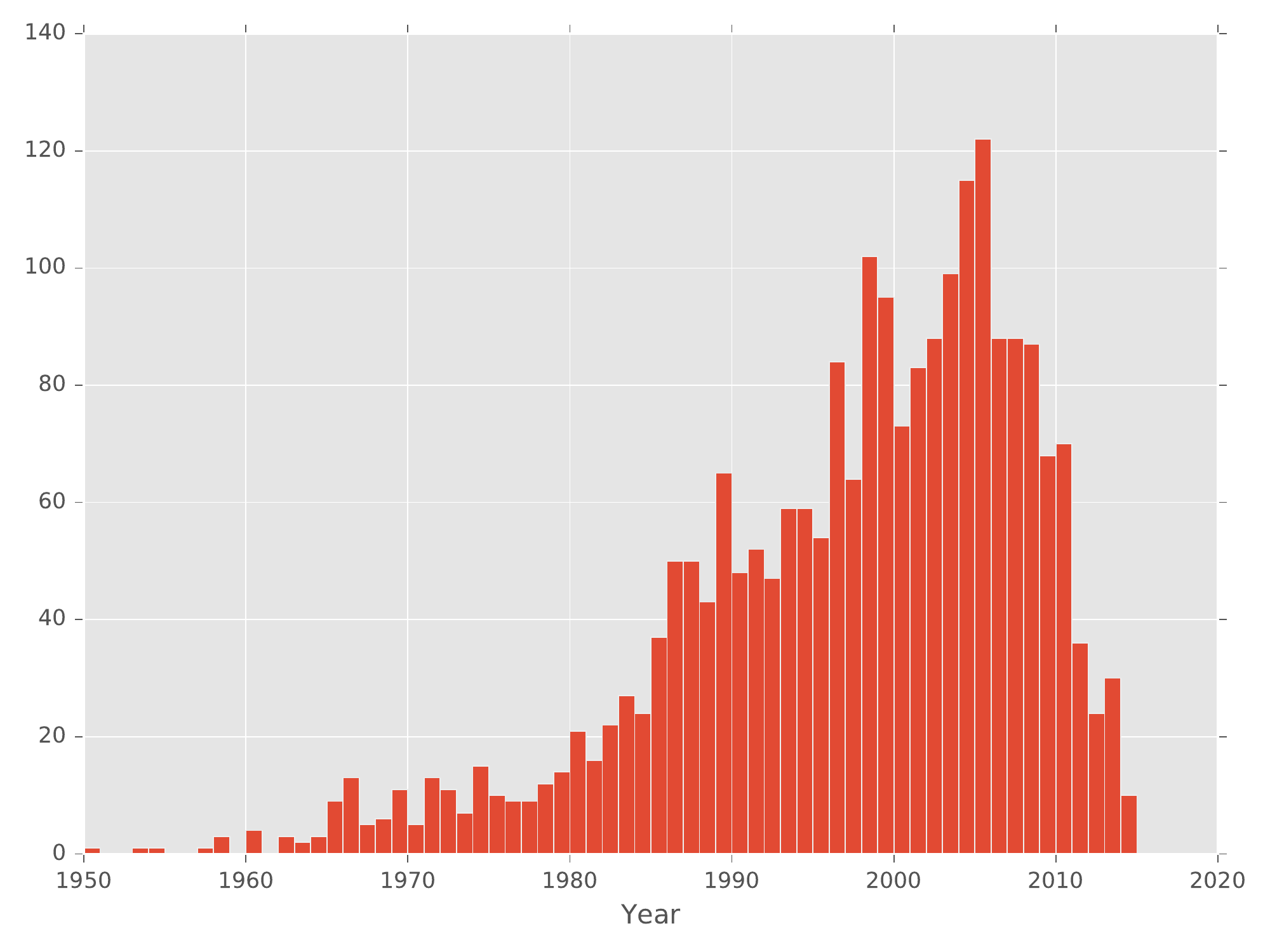}
\caption{Distribution of music groups by year of creation.\label{fig_sim}}
\end{figure}

Where possible, for each group we extracted its genre or genres (48\% of groups perform in one of the 275 genres), the list of performers, and the year of creation (only 50\% of groups have the year of establishment). Our data set covers 55 years: from 1960 to 2015. Fig.~\ref{fig_sim} shows the distribution of group creation years for the entire data set.

We arranged the discovered groups and musicians into a network. Each node in the network represents one music group. Two nodes are connected with an undirected arc if at least one musician participated in both groups (not necessarily at the same time). The resulting network is represented by an undirected, unweighted, simple graph. As expected, the graph is unconnected: there are minor ``islands'' of groups who did not connect with the ``mainland'' groups by sharing any performer known to us. The ``mainland''---the majority of all groups that can be reached by tracking shared performers---is called the giant connected component (GCC) of the graph. The GCC of our network contains 80\% of all recorded groups. All other connected components (the ``islands'') contain 13 or fewer groups each. This strong graph connectivity suggests that the phenomenon of sharing performers in the USSR and post-Soviet states was not unusual.

The constructed network has an excellent community structure. We define network community as a collection of groups that share more musicians among them than it is expected by chance~\cite{newman2003}. Modularity is the measure of the quality of community structure: the modularity of 1.0 represents perfectly separated communities, and a network with no community structure has the modularity of -0.5. Our network consists of 43 communities with the modularity of 0.76. The largest 25 communities contain 20 or more groups each. Latter analysis reveals that the partition of the network into communities is based on their dominant languages and genres.

For each node in the network (representing a single collective), we calculated six network measures: average neighbor degree; degree, closeness, betweenness, and eigenvalue centralities; and clustering coefficient. The degree of a node is a number of arcs incident to the node---in other words, the number of other groups that shared performers. The high degree represents high openness, mobility, dynamism, willingness to accept and donate human capital from and to other groups. The average neighbor degree looks at the same measure, but with respect to all network neighbors of the group. Sharing performers with a cohort of highly dynamic neighbors may be a decisive factor in success. The degree centrality is the degree normalized by the total number of possible connections (the size of the network minus 1).

The closeness centrality of a node is a measure of inverse proximity, in the number of arcs, from the node to all other nodes in the same network component. A high  degree centrality node is within few ``hops'' from the rest of the network; if sharing a performer between two groups is a proxy for mutual influence, then a higher degree centrality node is a stronger influencer and is subject to stronger influence than lower degree centrality nodes. 

A sequence of performer-sharing arcs that starts at one node and ends elsewhere is called a network path. There usually are many paths from one  group to another, and one or more of them is the shortest of all. The betweenness centrality of a node is a measure of how many shortest paths, altogether, pass through the node. In the same spirit as above, treating a sharing arc as a proxy for mutual influence, we can consider a path as a potential flow of influence, with longer paths being potentially weaker flows. Higher betweenness centrality means that a node belongs to many influence flows.

Eigenvalue centrality is the last centrality measure of our data set. It is defined recursively: the eigenvalue centrality of a node is the sum of its immediate neighbors' eigenvalue centralities.  

The clustering coefficient of a node is defined as the number of connections between the node's immediate neighbors divided by the total possible number of such connections. The clustering coefficient of 0 corresponds to a star network, where the central node is connected to its neighbors, but the neighbors are not connected to each other. The performers move from the group represented by the central node, to its neighbors, but not between the neighbors. The clustering coefficient of 1 corresponds to a complete graph, where all neighbors of the central node are also connected pairwise. There is at least one shared performer for every pair of the groups. 

We used all six network measures as independent predictor variables.

\subsection{Success Definition}

The success of modern post-Soviet and especially legacy Soviet-era popular music groups is particularly hard to quantify, in the first place due to the lack of proper music entertainment industry in the USSR and informal status of most of the groups, especially those performing rock and punk (\cite{troitsky1988,bright1985,fedorov1988}). Relatively few groups (mostly pop, disco, and folk) enjoyed official status and could afford to release and sell vinyl discs and high-quality magnetic tapes. However, the volumes of sales for them are not available. Billboard charts for Russian music groups and individual performers did not exist until the early 1990s. 

We propose to use secondary but easily collectible descriptors as proxies to the groups' success: combined visit frequencies of all group's Wikipedia pages (in all languages) recorded in 2011, 2013, and 2015, and the maximum current Google PageRank for all group's pages (in all languages). The visit frequency is the number of users who visit the page in one day, month or another time unit. The PageRank~\cite{ilprints422} is an integer number between 0 and 10 on the exponential scale that determines the relative importance of a web page. (Google itself has the top score of 10.)

Both measures---visit frequencies and PageRank---are easy to collect and hard to falsify. They are uniform with respect to the old (Soviet) and new (post-Soviet) groups, as well as groups performing in different genres, languages, and sociocultural contexts. They are naturally aligned with our source of structural data---the groups' Wikipedia pages. They essentially define success as long-term popularity: a group is successful if the information about it is or was actively sought at the moment of data collection, as well as two and four years ago. (We chose three sampling times to eliminate short-term spikes of interest that could have been caused by transient events.) 

To obtain the page visit statistics and Google PageRank, we used Wikipedia Article Traffic Statistics~\cite{wats} and PageRank calculator software~\cite{karaban2006}. We collected and recorded both measures simultaneously, over the period of several days, for 2,355 group pages, corresponding to 1,981 groups that were selected for further analysis. The Pearson correlation between the PageRank and the visit frequency for the selected pages is 0.29***, suggesting that they quantify different aspects of popularity.

According to our data, the most successful groups at the time of data collection (by the combination of both success proxies) were Tatu (ru; pop/pop-rock/electronic), O-Zone (ro; pop/electronic), Serebro (Silver; ru; europop), Nu Virgos (a.k.a. Via Gra; ua; dance-pop/europop), Aria (ru; heavy metal), Korol' i Shut (King and Jester; ru; punk), Leningrad (ru; punk/chanson), Lyapis Trubetskoy (by; rock/reggae/hip hop/alternative metal), Ranetki (ru; pop), Splean (ru; alternative rock/indie rock), Kino (Cinema; ru; punk/new wave/folk rock), Okean Elzy (Elza's Ocean; ua; rock), and Sektor Gaza (Gaza Strip; ru; rock/alternative metal).

\section{Results}
Since the goal of the study is to correlate the measures of success with the independent variables---the network measures---and possibly predict the measure of success, based on the same network measures, we applied two types of analysis to the data set: correlation analysis at the exploratory stage and machine learning (random decision forest) at the prediction stage.

\begin{table*}[!t]
\renewcommand{\arraystretch}{1.3}
\caption{Pearson correlations between six network measures and success proxies: Google PageRank, visit frequency, and logarithm of visit frequency\label{table1}}
\centering
\begin{tabular}{lrrr}
\hline
\bfseries Predictor feature&
\bfseries Google PageRank&
\bfseries Visit frequency&
\bfseries Log of visit frequency\\
\hline
Clustering coefficient & 0.018267 & -0.012364 & -0.010404\\
Eigenvalue centrality & 0.183748*** & 0.092697*** & 0.178790***\\
Betweenness centrality & 0.173226*** & 0.099439*** & 0.219597***\\
Average neighbors' degree & 0.125057*** & 0.056153 & 0.228551***\\
Degree centrality & 0.245450*** & 0.140752*** & 0.319802***\\
Closeness centrality & 0.193789*** & 0.121450*** & 0.339341***\\
\hline
\end{tabular}
\end{table*}

The linear regression shows that all network measures, except for the clustering coefficient, are significantly positively correlated with PageRank and the logarithm of the visit frequency (see Table~\ref{table1}). The Pearson correlation with the frequency itself is much weaker and, in the case of the average neighbors' degree, is not even significant.

\begin{figure*}[!tb]
\centering
\includegraphics[width=.65\textwidth]{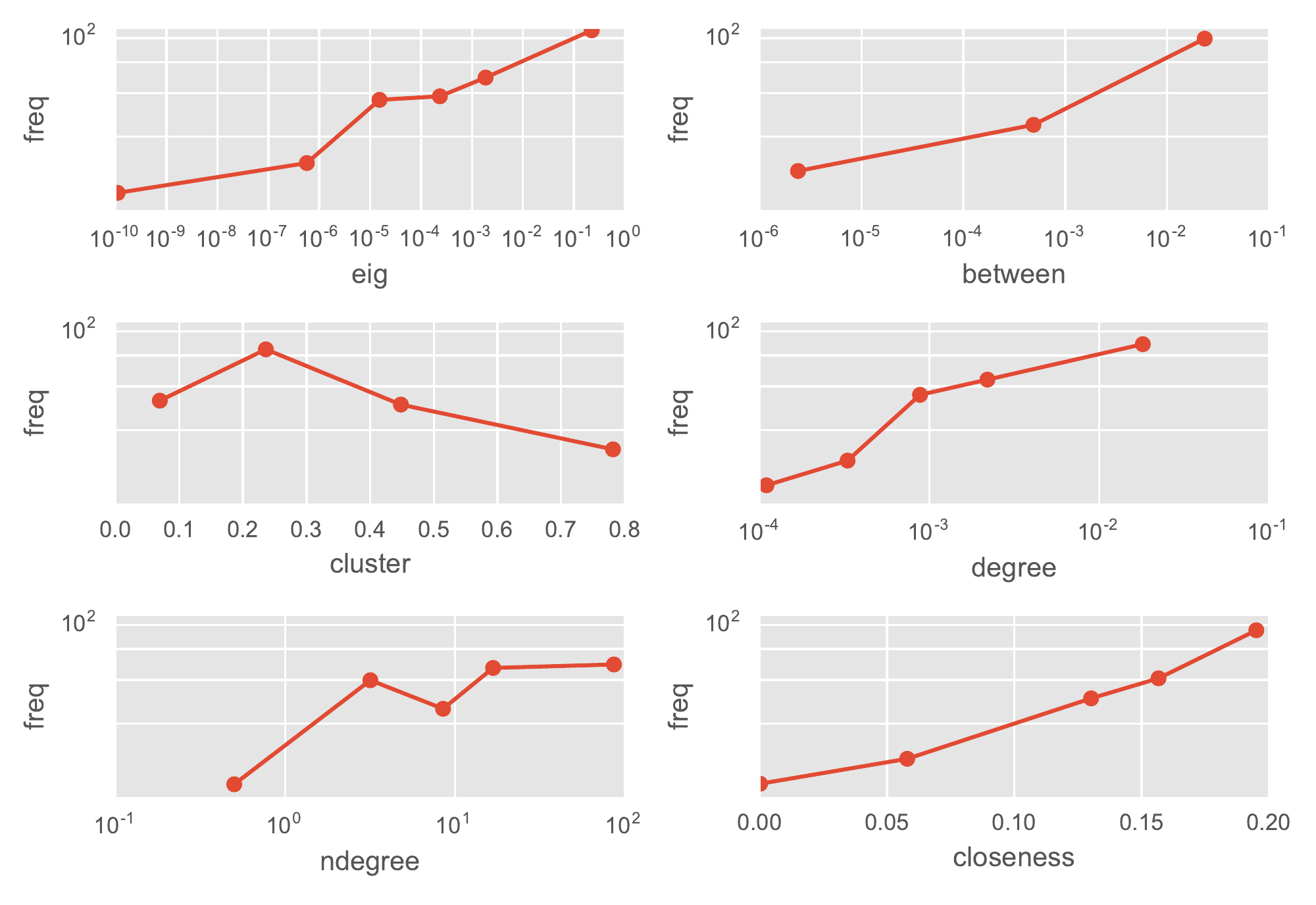}
\caption{Page visit frequency vs various network measures. Note that the horizontal axes for eigenvalue, betweenness, and closeness centralities, and the average neighbors' degree use logarithmic scale.\label{freq-trend}}
\end{figure*}

\begin{figure*}[!tb]
\centering
\includegraphics[width=.65\textwidth]{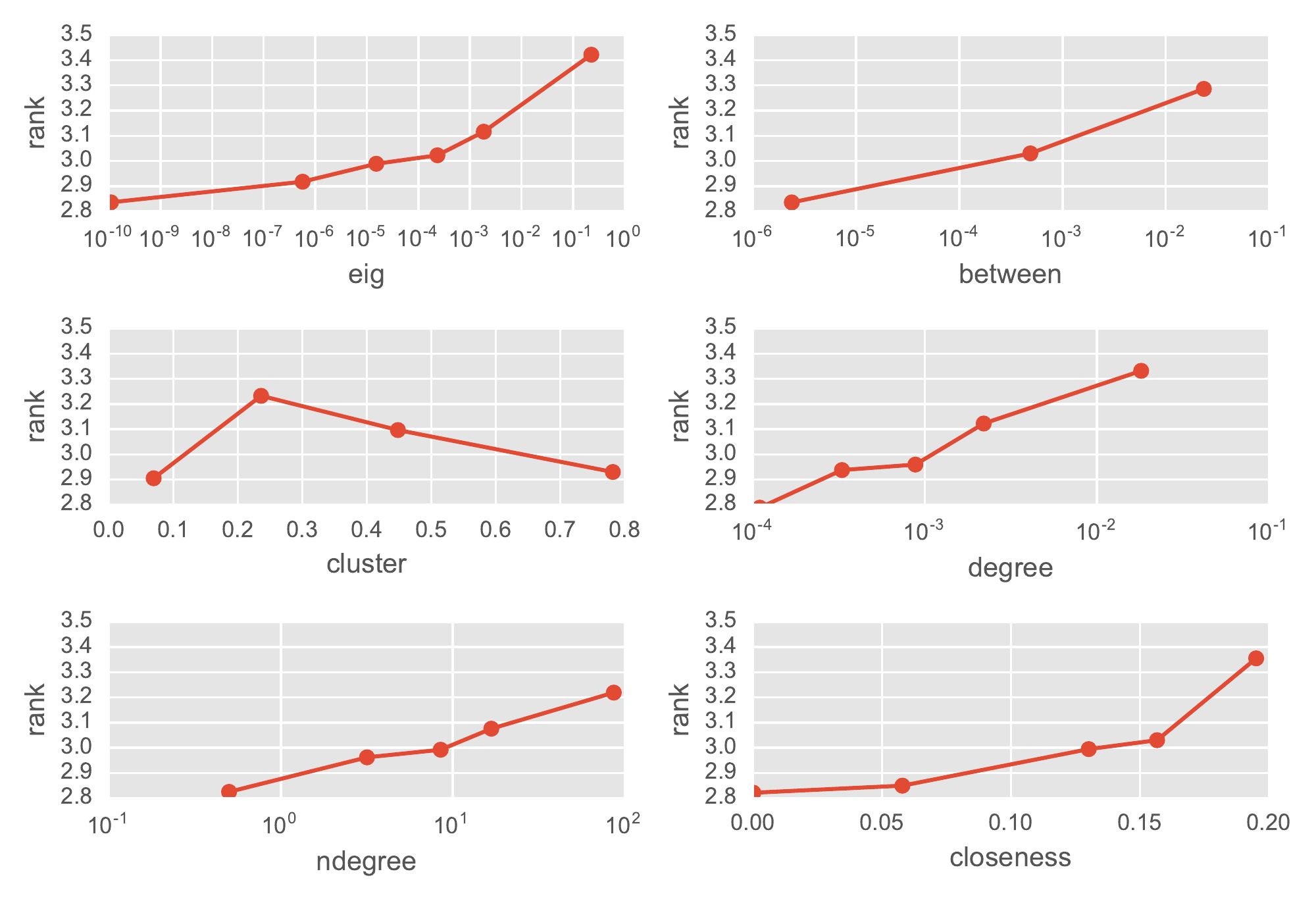}
\caption{Google PageRank vs various network measures. Note that the horizontal axes for eigenvalue, betweenness, and closeness centralities, and the average neighbors' degree use logarithmic scale.\label{rank-trend}}
\end{figure*}

For visualization, we split each independent variable range into several quantile ranges and plotted the mean value of each dependent variable versus the mean values of independent variables for each range (Figs.~\ref{freq-trend} and~\ref{rank-trend}). We can see from the figures that the near-zero Pearson correlations between the dependent variables and the clustering coefficient are explained by a highly nonlinear (parabolic) relationship between them.

\begin{table}[!b]
\renewcommand{\arraystretch}{1.3}
\caption{Scores of the random predictive forest models for success proxies, based on the network measures and group size\label{table2}}
\centering
\begin{tabular}{lrrrr}
\hline
&
\multicolumn{2}{c}{\bfseries Without group size}&
\multicolumn{2}{c}{\bfseries With group size}\\
\bfseries Success proxy&
\bfseries Training&
\bfseries Testing&
\bfseries Training&
\bfseries Testing
\\
\hline
PageRank&
0.877&
0.496&
0.895&
0.486\\
PageRank $\pm$1&
n/a&
0.895&
n/a&
0.927\\
Visit frequency (hi vs lo)&
0.910&
0.694&
0.925&
0.710\\
\hline
\end{tabular}
\end{table}

At the prediction stage, we attempted to predict the success proxies (Google PageRank and visit frequency), using the six network measures as predictors. The PageRank is restricted, by design, to eleven integer values (0 through 10) and can be treated as a categorical variable. The visit frequency, in general, is a continuous variable. However, for this study, we converted it into a categorical variable by separating the values above the median and below the median. This very coarse partitioning allowed us to obtain a much better prediction quality, which is essential for this pilot proof-of-concept study.

In the same spirit, we evaluated the quality of PageRank prediction two ways: by directly comparing the predicted and actual ranks and by allowing the difference of $\pm$1.

Aside from the six network measures that reflect performers' sharing, the Wikipedia PageRanks and visit frequencies are directly affected by the number of external links to the groups' pages. The numbers of links is positively correlated with group sizes (r=0.60***): if each performer has a Wikipedia page and a corpus of followers, then groups with more performers may have more links to them. Thus, the group size may partially or even fully explain the success proxies.  To control for the group size, we created two predictive models and added group size to one of them as yet another independent variable.

We used random decision forest classifier~\cite{breiman2001} to construct the predictive data models. We trained the models using 75\% randomly selected groups and tested the quality of fit on the remaining 25\% of the groups. Table~\ref{table2} has the summary of in-sample and out-of-sample scores (the fraction of correctly predicted ranks and visit frequencies). The most important feature in all models is the closeness centrality, which explains 20--25\% of the variance.

\section{Discussion}

The results presented in Tables~\ref{table1} and~\ref{table2} show that there is a weak, but statistically significant relationship between the centrality and clustering measures in the network of performers' transfers among non-academic Soviet and post-Soviet groups, and the proxies for groups' success, such as Google PageRank and visit frequency of the groups' Wikipedia pages. This relationship can be used both to explain groups' popularity and predict it, based solely on the network topology.

One of the features that is most positively correlated with the success proxies, the degree centrality (Table~\ref{table1}), represents the number of immediate network connections and is directly related to the number of cooperations with other groups (each connection stands for one or more shared performers). We hypothesize that music groups benefit from the cultural cross-pollination caused by performers moving between different projects.

The closeness centrality, the other most positively correlated variable, is a measure of distance from the group to all other groups in the same network component, expressed via performers' sharing. Considering performers' migration in and out of a group as an act of cultural influence~\cite{mauskapf2016}, we further hypothesize that if a group is located in close proximity of the majority of the other groups, it is either subject to higher level of influence by the other groups or exercises higher influence onto the other groups. In other words, it either attracts musicians who bring fresh ideas or provides musicians who can be traced back to the group and enhance its popularity.

The remaining three centrality measures: average neighbors' degree, betweenness centrality, and eigenvalue centrality---are still positively correlated with the success proxies, but to a lesser extent. 

If the degree centrality of a node is correlated with its success, then the average degree centrality of a group's neighbors is correlated with their average success. It is tempting to conclude that the average success of a group's neighbors is also similar to the group's success. The measure of such similarity is called assortativity~\cite{newman2006} and is, essentially, a correlation between the attributes of network neighbors. We found that the assortativity of our network is 0.064 regarding PageRank and 0.027 regarding visit frequency. Both assortativities are marginally positive and suggest that the average neighbors' degree affects the success proxies of a node only indirectly. At the moment, we do not have an explanation of this mechanism.

The eigenvalue centrality recursively defines the ``importance'' of a group through the ``importance'' of its neighbors. The influence mechanism of the eigenvalue centrality is not clear to us so far as well.

The clustering coefficient is the only feature that has a non-linear---parabolic---relationship with the success proxies (Figs.~\ref{freq-trend} and~\ref{rank-trend}). The proxies are lower for low and high values of the coefficient, and reach the maximum at 0.2--0.3. In other words, a group is most successful when the density of the surrounding network of neighbors is between 20\% and 30\%. This observation can be explained in terms of fold networks, as described in~\cite{vaan2015} for the teams of video game developers: ``Teams are most likely to produce games that stand out and are recognized as outstanding when their cognitively heterogeneous communities have points of intersection''---but do not fully intersect. In the case of no intersection, the clustering coefficient is 0; in the case of full intersection, it is 1. The optimal intermediate value of the coefficient that we observed in the network of music groups is consistent with the theory of fold networks. The same parabolic behavior has been observed in a network of artists by Giuffre~\cite{giuffre1999} and the network of Broadway musicians by Uzzi and Spiro~\cite{uzzi2005}.

The results of the predictive study confirm our expectations that the coarse success metrics can be predicted better than by chance based only on the topological characteristics of the musician sharing network. We may obtain better predictions in the future by considering directed, weighted network graphs that describe directions and intensity of performers' sharing.

\section{Conclusion}
In this paper, we explored the relationship between success proxies of Soviet and post-Soviet non-academic music groups and the patterns of performers' sharing between the groups. We used Google PageRank and visit frequency of the groups' Wikipedia pages as success proxies (dependent variables), and a variety of network measures (degrees, centralities, and clustering coefficient) as independent variables. We detected weak, but statistically significant correlations between degrees and centralities, on the one hand, and the success proxies, on the other hand, that suggest that a group's success is positively influenced by the number of other groups with shared performers. The most successful groups also had an intermediate (not too low and not too high) clustering coefficient, making the group's network neighbors moderately, but not fully, interconnected. Such interconnection pattern supports mutual influence but also preserves groups' identity.

We used the same set of independent variables and machine learning algorithms to predict groups' success proxies. The results of the random decision forest-based predictions are substantially better than by chance.

We believe that our network-based success exploration and prediction approach can be easily extended to other areas of arts, humanities, sciences, and business that have medium- to long-term team-based collaborations.

\section*{Acknowledgment}
We would like to thank Yongren Shi (Cornell University) and the attendees of the International Conference on Computational Social Science--2016 for their inspiring questions and valuable feedback.

\bibliographystyle{IEEEtran}
\bibliography{cs}

\begin{thebibliography}{10}
\providecommand{\url}[1]{#1}
\csname url@samestyle\endcsname
\providecommand{\newblock}{\relax}
\providecommand{\bibinfo}[2]{#2}
\providecommand{\BIBentrySTDinterwordspacing}{\spaceskip=0pt\relax}
\providecommand{\BIBentryALTinterwordstretchfactor}{4}
\providecommand{\BIBentryALTinterwordspacing}{\spaceskip=\fontdimen2\font plus
\BIBentryALTinterwordstretchfactor\fontdimen3\font minus
  \fontdimen4\font\relax}
\providecommand{\BIBforeignlanguage}[2]{{%
\expandafter\ifx\csname l@#1\endcsname\relax
\typeout{** WARNING: IEEEtran.bst: No hyphenation pattern has been}%
\typeout{** loaded for the language `#1'. Using the pattern for}%
\typeout{** the default language instead.}%
\else
\language=\csname l@#1\endcsname
\fi
#2}}
\providecommand{\BIBdecl}{\relax}
\BIBdecl

\bibitem{ghasemian2016}
F.~Ghasemian, K.~Zamanifar, N.~Ghasem-Aqaee, and N.~Contractor, ``Toward a
  better scientific collaboration success prediction model through the feature
  space expansion,'' \emph{Scientometrics}, vol. 108, no.~2, pp. 777--801,
  2016.

\bibitem{vaan2015}
M.~de~Vaan, D.~Stark, and B.~Vedres, ``Game changer: The topology of
  creativity,'' \emph{American Journal of Sociology}, vol. 120, no.~4, pp.
  1144--1194, 2015.

\bibitem{giuffre1999}
K.~Giuffre, ``Sandpiles of opportunity: Success in the art world,''
  \emph{Social Forces}, vol.~77, no.~3, pp. 815--832, 1999.

\bibitem{preece2001}
J.~Preece, ``Sociability and usability in online communities: determining and
  measuring success,'' \emph{Behaviour and Information Technology}, vol.~20,
  no.~5, pp. 347--356, 2001.

\bibitem{hirschman1985}
E.~Hirschman and A.~Pieros, ``Relationships among indicators of success in
  broadway plays and motion pictures,'' \emph{Journal of Cultural Economics},
  vol.~9, no.~1, pp. 35--63, 1985.

\bibitem{hadida2010}
A.~Hadida, ``Commercial success and artistic recognition of motion picture
  projects,'' \emph{Journal of Cultural Economics}, vol.~34, no.~1, pp. 45--80,
  2010.

\bibitem{uzzi2005}
B.~Uzzi and J.~Spiro, ``Collaboration and creativity: The small world
  problem,'' \emph{American Journal of Sociology}, vol. 111, no.~2, pp.
  447--504, 2005.

\bibitem{mauskapf2016}
M.~Mauskapf, E.-A. Horvat, N.~Askin, and B.~Uzzi, ``Understanding the link
  between quality, social influence, and success in popular music,'' 2016,
  international Conference on Computational Social Science, Evanston, IL
  (working paper).

\bibitem{granovetter1973}
M.~Granovetter, ``The strength of weak ties,'' \emph{American Journal of
  Sociology}, vol.~78, no.~6, pp. 1360--1380, May 1973.

\bibitem{troitsky1988}
A.~Troitsky, \emph{Back in the {USSR}: The true story of rock in Russia}.\hskip
  1em plus 0.5em minus 0.4em\relax London and Boston: Faber and Faber, 1988.

\bibitem{pilkington2001}
H.~Pilkington, ``Reconfiguring ``the {W}est'','' in \emph{Looking West?
  Cultural Globalization and Russian Youth Cultures}.\hskip 1em plus 0.5em
  minus 0.4em\relax University Park, PA: The Pennsylvania University Press,
  2001, pp. 165--200.

\bibitem{bright1986}
T.~Bright, ``Pop music in the {USSR},'' \emph{Media, Culture and Society},
  vol.~8, no.~3, pp. 357--369, 1986.

\bibitem{partan2007}
O.~Partan, ``Alla: The jester-queen of {R}ussian pop culture,'' \emph{The
  Russian Review}, vol.~66, no.~3, pp. 483--500, 2007.

\bibitem{steinholt2003}
Y.~Steinholt, ``You can't rid a song of its words: notes on the hegemony of
  lyrics in {Russian} rock songs,'' \emph{Popular Music}, vol.~22, no.~1, pp.
  89--108, 2003.

\bibitem{roy2010}
W.~Roy and T.~Dowd, ``What is sociological about music?'' \emph{Annual Review
  of Sociology}, vol.~35, pp. 183--203, 2010.

\bibitem{krackhardt1998}
D.~Krackhardt and K.~Carley, ``A {PCANS} model of structure in organizations,''
  in \emph{Proc. 1998 International Symposium on Command and Control Research
  and Technology}, Monterey, CA, 1998.

\bibitem{park2015}
D.~Park, A.~Bae, M.~Schich, and J.~Park, ``Topology and evolution of the
  network of {W}estern classical music composers,'' \emph{EPJ Data Science},
  vol.~4, no.~2, 2015, doi: 10.1140/epjds/s13688-015-0039-z.

\bibitem{mauskapf}
M.~Mauskapf and N.~Askin, ``What makes popular culture popular?: Cultural
  networks and optimal differentiation in music,'' {A}merican Sociological
  Review (submitted).

\bibitem{newman2003}
M.~Newman, ``Mixing patterns in networks,'' \emph{Physical Review E}, vol.~67,
  2003.

\bibitem{bright1985}
T.~Bright, ``Soviet crusade against pop,'' \emph{Popular Music}, no.~5, pp.
  123--148, 1985.

\bibitem{fedorov1988}
E.~Fedorov, \emph{Rok v neskol'kikh litsakh [Rock in several faces]}.\hskip 1em
  plus 0.5em minus 0.4em\relax Moscow: Molodaia Gvardiia, 1988.

\bibitem{ilprints422}
\BIBentryALTinterwordspacing
L.~Page, S.~Brin, R.~Motwani, and T.~Winograd, ``{The PageRank Citation
  Ranking: Bringing Order to the Web},'' Stanford InfoLab, Technical Report
  1999-66, November 1999, previous number = SIDL-WP-1999-0120. [Online].
  Available: \url{http://ilpubs.stanford.edu:8090/422}
\BIBentrySTDinterwordspacing

\bibitem{wats}
``Wikipedia article traffic statistics,'' 2016, retrieved from
  http://stats.grok.se.

\bibitem{karaban2006}
Y.~Karaban, ``{WWW::Google::PageRank (Version 0.19) [Software]},'' 2006,
  available from
  http://search.cpan.org/\~{}ykar/WWW-Google-PageRank/lib/WWW/Google/PageRank.%
pm.

\bibitem{breiman2001}
L.~Breiman, ``Random forests,'' \emph{Machine Learning}, vol.~45, no.~1, pp.
  5--32, 2001.

\bibitem{newman2006}
M.~Newman, ``Modularity and community structure in networks,''
  \emph{Proceedings of the National Academy of Sciences of the United States of
  America}, vol. 103, no.~23, pp. 8577--8696, 2006.

\end{thebibliography}

\end{document}